# Scattering theory with a natural regularization: Rediscovering the J-matrix method

A. D. Alhaidari[a,+], H. Bahlouli[b], M. S. Abdelmonem[b],
F. S. Al-Ameen[c] and T. H. Al-Abdulaal[c]

[a] *Shura Council, Riyadh 11212, Saudi Arabia*
[b] *Physics Department, King Fahd University of Petroleum & Minerals, Dhahran 31261, Saudi Arabia*
[c] *Girls College of Sciences, Dammam 31113, Saudi Arabia*

In three dimensional scattering, the energy continuum wavefunction is obtained by utilizing two independent solutions of the reference wave equation. One of them is typically singular (usually, near the origin of configuration space). Both are asymptotically regular and sinusoidal with a phase difference (shift) that contains information about the scattering potential. Therefore, both solutions are essential for scattering calculations. Various regularization techniques were developed to handle the singular solution leading to different well-established scattering methods. To simplify the calculation the regularized solutions are usually constructed in a space that diagonalizes the reference Hamiltonian. In this work, we start by proposing solutions that are already regular. We write them as infinite series of square integrable basis functions that are compatible with the domain of the reference Hamiltonian. However, we relax the diagonal constraint on the representation by requiring that the basis supports an infinite tridiagonal matrix representation of the wave operator. The hope is that by relaxing this constraint on the solution space a larger freedom is achieved in regularization such that a natural choice emerges as a result. We find that one of the resulting two independent wavefunctions is, in fact, the regular solution of the reference problem. The other is uniquely regularized in the sense that it solves the reference wave equation only outside a dense region covering the singularity in configuration space. However, asymptotically it is identical to the irregular solution. We show that this natural and special regularization is equivalent to that already used in the J-matrix method of scattering.

PACS numbers: 03.65.Fd, 03.65.Nk, 03.65.Ca

Keywords: scattering, regularization, tridiagonal representation, recursion relation, J-matrix method.

## I. INTRODUCTION

One of the most fundamental problems in theoretical physics is the solution of the energy eigenvalue problem $H|\psi\rangle = E|\psi\rangle$, where $H$ is the Hamiltonian of the system, $E$ is the energy and $\psi$ is the wavefunction. The solution of this equation (i.e., the wavefunction and corresponding energy spectrum) contains all information about the structure and dynamics of the system. The energy spectrum is either discrete (for bound states), continuous (for scattering states), or consists of a set of energy bands with forbidden gaps in between [1]. Additionally, investigation of scattering could be enhanced by the analytic extension to the complex energy plane which makes the study of the resonance structure easier and more manageable [2]. The solution of this equation for a general observable $H$

---

[+] Corresponding Author, email: haidari@mailaps.org

is often very difficult to obtain. However, for simple systems or for those with high degree of symmetry, an analytic solution is feasible. On the other hand, for a large class of problems that model realistic physical systems, the Hamiltonian could be written as the sum of two components: $H = H_0 + V$. The "reference Hamiltonian" $H_0$ is often simpler and carries a high degree of symmetry. However, its contribution which includes the kinetic energy operator and possibly a reference potential $V_0$ (e.g., the Coulomb), extends all the way to infinity at the boundary of configuration space. On the other hand, the scattering potential $V$ does not have the same high degree of symmetry as that of $H_0$. However, it is usually endowed with either one of two properties. Its contribution is either very small compared to $H_0$ or is limited to a finite region in configuration or function space. Perturbation techniques are used to give a numerical evaluation of its contribution in the former case, whereas algebraic methods, such as the R-matrix [3] and J-matrix [4], are used in the latter. Thus, the analytic problem is confined to finding solutions of the reference $H_0$–problem, $H_0 |\chi\rangle = E |\chi\rangle$.

Typically, the realization of $H_0$ in configuration space is given by a second order differential operator. Therefore, the reference wave equation has two independent solutions. However, one of these two solutions is singular (irregular); typically, in a dense region around the origin of configuration space. Both behave asymptotically (far away from the scattering region where the potentials vanish) as free particles. That is, they have sinusoidal behavior as sine-like and cosine-like solutions. These two solutions are essential in scattering calculations where they are augmented by the contribution of the scattering potential $V$ to give the phase shift [5]. To carry out the calculations, where one builds on these reference solutions, one needs to manipulate regular objects with finite values. Consequently, several algebraic and analytic regularization techniques were developed to deal with the singular solution so that one could extract meaningful scattering information. This has been accomplished in various methods with varying level of accuracy and degree of convergence. For simplicity of the calculations one traditionally works in a basis that carries a diagonal matrix representation of the reference Hamiltonian. In this work, we remove the diagonal constraint on the representation by searching for a discrete square integrable basis that supports an infinite tridiagonal matrix representation of $H_0$. Doing so will, of course, make the solution space regular from the start. Nonetheless, our hope is that by relaxing the diagonal constraint on the solution space a larger freedom is achieved in regularization such that a *natural* choice could emerge as a result. Now, due to the higher degree of symmetry of the reference problem, it is sometimes possible to find such a special basis, $\{\phi_n\}_{n=0}^{\infty}$. That is, the matrix representation of the reference wave operator $H_0 - E$ in this basis is tridiagonal for all $E$. Precisely, $\langle \phi_n | H_0 - E | \phi_m \rangle = J_{nm}(E)$ such that $J_{nm} = 0$ for $|n - m| > 1$, where $n, m = 0, 1, 2, ...$ This will allow for an algebraic solution of the eigenvalue wave equation with a continuous value of the energy; a property which is desirable for scattering. On the other hand, the diagonal representation can only admit discrete eigenenergies which are compatible with the bound states. Additionally, solutions of the reference $H_0$–problem will be obtained algebraically in terms of orthogonal polynomials that satisfy the three-term recursion relation resulting from the tridiagonal structure of the matrix representation of the reference wave equation. We show that the regular solution of the reference problem (e.g., the one that behaves asymptotically as sine-like, $\chi_{\sin}$) could be written as an infinite sum of the basis elements with the orthogonal polynomials in the expansion coefficients. That is, we can write $\chi_{\sin}(\vec{r},E) = \sum_n s_n(\mu)\phi_n(\vec{r})$, where $\mu$ is a real energy-dependent parameter and the reference wave equation is equivalent to the matrix equation $\sum_m J_{nm} s_m = \sum_{m=n, n\pm 1} J_{nm} s_m = 0$.

Subsequently, we show that the sine-like expansion coefficients, $\{s_n\}_{n=0}^{\infty}$, satisfy a second order linear differential equation in the energy parameter $\mu$. Hence, we find another independent set of solutions to this equation; say $\{c_n(\mu)\}_{n=0}^{\infty}$. However, we will discover that these expansion coefficients satisfy the same three-term recursion relation as $\{s_n\}$ except for the initial relation ($n = 0$). That is, $\sum_m J_{nm} c_m = 0$ for all $n \neq 0$. Precisely, $\sum_m J_{nm} c_m = \alpha(E)\delta_{n0}$, where $\alpha$ is real and energy dependent. Therefore, the corresponding wavefunction, $\chi_{\cos}(\vec{r},E) = \sum_n c_n(\mu)\phi_n(\vec{r})$, does not satisfy the reference wave equation. It satisfies a regularized non-homogeneous wave equation that reads

$$(H_0 - E)|\chi_{\cos}\rangle = \alpha(E)|\tilde{\phi}_0\rangle, \qquad (1.1)$$

where $\tilde{\phi}_0$ is an element of the set $\{\tilde{\phi}_n\}_{n=0}^{\infty}$ which is orthogonal to $\{\phi_n\}_{n=0}^{\infty}$ (i.e., $\langle\phi_n|\tilde{\phi}_m\rangle = \langle\tilde{\phi}_n|\phi_m\rangle = \delta_{nm}$). However, asymptotically $\chi_{\cos}(r,E)$ is sinusoidal (cosine-like) where it is identical to the irregular solution of the reference problem. Therefore, the singular solution becomes regularized in the sense that it solves the modified wave equation (1.1).

In the following section we consider the three dimensional problem with spherical symmetry where the reference Hamiltonian is the partial $\ell$-wave free Hamiltonian. We propose an $L^2$ basis compatible with the domain of $H_0$ and find two independent set of expansion coefficients for the reference wavefunction. We show that one of these two wavefunctions is the exact regular solution of the reference problem. The other one is a uniquely regularized version of the irregular solution. We also show that the exact irregular solution and the "regularized" wavefunction are asymptotically equal. In Sec. III, the resulting regularization will be compared to that which is already used in the J-matrix method of scattering [5].

## II. REGULARIZATION OF THE REFERENCE SOLUTION

The time-independent radial Schrödinger equation for a scalar particle in the field of a central potential $V(r)$ reads as follows

$$\left[-\frac{1}{2}\frac{d^2}{dr^2} + \frac{\ell(\ell+1)}{2r^2} + V(r) - E\right]\psi_\ell(r,E) = 0, \qquad (2.1)$$

where $\ell$ is the angular momentum quantum number and we have used the atomic units $\hbar = m = 1$. Now, we assume that the range of the potential is finite and thus take the reference Hamiltonian $H_0$ to be the free kinetic energy operator, $H_0 = -\frac{1}{2}\frac{d^2}{dr^2} + \frac{\ell(\ell+1)}{2r}$. Therefore, the reference wave equation becomes

$$\left[-\frac{d^2}{dr^2} + \frac{\ell(\ell+1)}{r^2} - 2E\right]\chi_\ell(r,E) = 0. \qquad (2.2)$$

The two independent scattering solutions (for $E > 0$) of this equation which are also energy eigenfunctions of $H_0$, could be found in most standard textbooks on quantum mechanics [6]. They are written in terms of the spherical Bessel and Neumann functions as follows:

$$\chi_{reg}(r,E) = \tfrac{2}{\sqrt{\pi}}(kr) j_\ell(kr), \qquad (2.3a)$$

$$\chi_{irr}(r,E) = \tfrac{2}{\sqrt{\pi}}(kr) n_\ell(kr), \qquad (2.3b)$$

where $k = \sqrt{2E}$. The regular solution is energy-normalized, $\langle \chi_{reg} | \chi'_{reg} \rangle = \delta(k-k')$, whereas the irregular solution is not square integrable (with respect to the integration measure, $dr$). Near the origin they behave as $\chi_{reg} \to r^{\ell+1}$ and $\chi_{irr} \to r^{-\ell}$. On the other hand, asymptotically ($r \to \infty$) they are sinusoidal: $\chi_{reg} \to \tfrac{2}{\sqrt{\pi}} \sin(kr - \pi\ell/2)$ and $\chi_{irr} \to -\tfrac{2}{\sqrt{\pi}} \cos(kr - \pi\ell/2)$. Now, we search for a complete $L^2$ basis functions, $\{\phi_n\}_{n=0}^{\infty}$, for $\chi_{reg}$ that could also support an infinite tridiagonal matrix representation for the reference wave operator $H_0 - E$. One such basis which is compatible with $\chi_{reg}$ (i.e., defined in the same range $r \in [0,\infty]$, behaves at the origin as $r^{\ell+1}$, and square integrable) is

$$\phi_n(r) = (\lambda r)^{\ell+1} e^{-\lambda r/2} L_n^\nu(\lambda r), \qquad (2.4)$$

where $L_n^\nu(x)$ is the associated Laguerre polynomial of order $n$, $\nu > -1$, and $\lambda$ is a positive basis parameter of inverse length dimension. Using the differential equation of the Laguerre polynomials [7] and their differential formula, $x\tfrac{d}{dx} L_n^\lambda = n L_n^\lambda - (n+\lambda) L_{n-1}^\lambda$, we can write

$$(H_0 - E)|\phi_n\rangle = \left[ \tfrac{n}{2r}\left(\lambda + \tfrac{\nu - 2\ell - 1}{r}\right) + \lambda \tfrac{\ell+1}{2r} - \tfrac{\lambda^2}{8} - E \right] |\phi_n\rangle$$
$$+ \tfrac{n+\nu}{2r^2}(2\ell+1-\nu) |\phi_{n-1}\rangle \qquad (2.5)$$

If we project on the left by $\langle \phi_m |$ then the orthogonality relation for the Laguerre polynomials [7] dictates that a tridiagonal representation is obtained only if $\nu = 2\ell+1$. Moreover, using the recursion relation of the Laguerre polynomials and their orthogonality property we obtain the following tridiagonal representation of the reference wave operator

$$\langle \phi_n | H_0 - E | \phi_m \rangle = J_{nm}(E) = \tfrac{\Gamma(n+2\ell+2)}{\lambda \Gamma(n+1)}(E + \lambda^2/8) \times$$
$$\left[ -2(n+\ell+1) \tfrac{E - \lambda^2/8}{E + \lambda^2/8} \delta_{n,m} + n \delta_{n,m+1} + (n+2\ell+2) \delta_{n,m-1} \right] \qquad (2.6)$$

Therefore, if we write

$$\chi_{\sin}(r,E) \equiv \chi_{reg} = \sum_{n=0}^{\infty} s_n(E) \phi_n(r), \qquad (2.7)$$

Then the sine-like expansion coefficients, $\{s_n\}_{n=0}^{\infty}$, satisfy the three term recursion relation obtained from (2.6) as $\sum_m J_{nm} s_m = 0$, which reads

$$2(n+\ell+1) y s_n = n s_{n-1} + (n+2\ell+2) s_{n+1}, \qquad (2.8)$$

where $y = \tfrac{E - \lambda^2/8}{E + \lambda^2/8}$. Rewriting this recursion relation in terms of the polynomials $P_n(E) = [\Gamma(n+2\ell+2)/\Gamma(n+1)] s_n(E)$, we obtain the more familiar recursion relation

$$2(n+\ell+1) y P_n = (n+2\ell+1) P_{n-1} + (n+1) P_{n+1}, \; n = 1, 2, \ldots, \qquad (2.9a)$$

$$2(\ell+1) y P_0 = P_1, \qquad (2.9b)$$

which is that of the Gegenbauer polynomial $C_n^{\ell+1}(y)$ [7]. Thus, $\{s_n\}$ can now be determined modulo an arbitrary real function of the energy, which is independent of the index $n$. However in this work we will not use the recursion relation to obtain $\{s_n\}$ but instead we evaluate $\{s_n\}$ using the orthogonality property of the Laguerre polynomials in (2.7) as

$$s_n(E) = \frac{\Gamma(n+1)}{\Gamma(n+2\ell+2)} \int_0^\infty x^\ell e^{-x/2} L_n^{2\ell+1}(x) \chi_{reg}(x/\lambda, E) \, dx, \qquad (2.10)$$

where $x = \lambda r$. Rewriting the wavefunction (2.3a) in terms of the Bessel function $J_{\ell+\frac{1}{2}}(z)$ $= \sqrt{\frac{2z}{\pi}} j_\ell(z)$ and substituting in (2.10) we obtain

$$s_n(E) = \sqrt{2\mu} \, \frac{\Gamma(n+1)}{\Gamma(n+2\ell+2)} \int_0^\infty x^{\ell+\frac{1}{2}} e^{-x/2} L_n^{2\ell+1}(x) J_{\ell+\frac{1}{2}}(\mu x) \, dx, \qquad (2.11)$$

where $\mu = k/\lambda$. This integral is not found in mathematical tables but has recently been evaluated by one of the authors in [8]. The result is

$$s_n(E) = \frac{1}{\sqrt{\pi}} 2^{\ell+1} \frac{\Gamma(n+1)\Gamma(\ell+1)}{\Gamma(n+2\ell+2)} (\sin\theta)^{\ell+1} C_n^{\ell+1}(\cos\theta), \qquad (2.12)$$

where $\cos\theta = \frac{\mu^2 - 1/4}{\mu^2 + 1/4} = y$, and $0 < \theta \leq \pi$. Using the differential equation for $C_n^\nu(y)$, one can show that $s_n(E)$ satisfies the following second order differential equation

$$\left[(1-y^2)\frac{d^2}{dy^2} - y\frac{d}{dy} - \frac{\ell(\ell+1)}{1-y^2} + (n+\ell+1)^2\right] s_n(E) = 0, \qquad (2.13)$$

We can find a second independent solution for this equation. Let's call it $c_n(E)$ and using the fact that $y = \pm 1$ are regular singular points then Frobenius method dictates that the solution has the following form [9]

$$c_n(E) = (1-y)^\alpha (1+y)^\beta f_n(\alpha, \beta; y), \qquad (2.14)$$

where $\alpha$ and $\beta$ are real parameters such that $\beta > 0$ to prevent infrared divergence (at $E = 0$ where $y = -1$). It should, thus, be obvious that the solution that simultaneously satisfies the recursion relation (2.8) and the differential equation (2.13) will be unique modulo an arbitrary normalization factor which is independent of $E$ and $n$. That is, it will only depend on the angular momentum $\ell$ and we refer to it as $A_\ell$. Substituting (2.14) in place of $s_n(E)$ in Eq. (2.13) shows that $f_n(\alpha, \beta; y)$ satisfies the same differential equation as the hypergeometric function ${}_2F_1\left(a, b; c; \frac{1-y}{2}\right)$ [7] provided that

(1) $c = 2\alpha + \frac{1}{2}$, $a = \alpha + \beta \pm (n+\ell+1)$, $b = \alpha + \beta \mp (n+\ell+1)$, and   (2.15a)

(2) $\alpha(2\alpha - 1) + \beta(2\beta - 1) = \ell(\ell+1)$   (2.15b)

Additionally, we must impose the condition that either $\alpha = \beta$ or $\alpha + \beta = \frac{1}{2}$. Next, we will investigate these two cases separately. We refer to the resulting expansion coefficients that satisfy the recursion relation (2.8) for all $n$ by $s_n(E)$. Others that also do, but only for $n \neq 0$, will be referred to as $c_n(E)$. Thus, the wavefunction (2.7) with the expansion coefficients $\{s_n\}$ satisfy the reference wave equation $(H_0 - E)|\chi\rangle = 0$ whereas that with $\{c_n\}$ does not. Nonetheless, the latter will be considered as the regularized version of the irregular solution in the sense of regularization defined in the Introduction.

### A. The case $\alpha = \beta$

Maintaining positivity of $\beta$, this case produces two solutions. One is valid only for S-wave ($\ell = 0$) whereas the other is true for all values of the angular momentum $\ell$. Hence, there will be two inequivalent solutions for $\ell = 0$. For general $\ell$, Eq. (2.15) gives

$$\alpha = \beta = \tfrac{1}{2}(\ell+1),\ a = -n,\ b = n+2\ell+2,\ \text{and}\ c = \ell + \tfrac{3}{2} \tag{2.16}$$

This is the regular solution (2.3a) which we have already found in (2.7) and (2.12) and called it $s_n(E)$. This could easily be seen by noting that $_2F_1\left(-n, n+2\ell+2; \ell+\tfrac{3}{2}; \tfrac{1-y}{2}\right)$ is proportional to $C_n^{\ell+1}(y)$ [10] whereas $(1-y)^\alpha(1+y)^\beta = \left(1-y^2\right)^{\frac{\ell+1}{2}} = (\sin\theta)^{\ell+1}$. However, for S-wave ($\ell = 0$) there exists another independent special solution where,

$$\alpha = \beta = -\tfrac{1}{2}\ell = 0,\ a = -b = -n-1,\ \text{and}\ c = \tfrac{1}{2}, \tag{2.17}$$

corresponding to $_2F_1\left(-n-1, n+1; \tfrac{1}{2}; \tfrac{1-y}{2}\right)$, which is the Chebyshev polynomial of the first kind, $T_{n+1}(y)$ [7]. Therefore, the expansion coefficients of the reference wavefunction in (2.7) are

$$c_n(E) = \tfrac{A}{n+1} T_{n+1}(\cos\theta) = \tfrac{A}{n+1}\cos(n+1)\theta, \tag{2.18}$$

where $A$ is a normalization constant, which is independent of the energy $E$ and the index $n$. Now, this solution satisfies the three term recursion relation (2.8) with $\ell = 0$, but not the initial relation (i.e., for $n = 0$). That's why we called it $c_n(E)$ and not $s_n(E)$. In fact, one can easily show that it satisfies an inhomogeneous initial relation which reads as follows

$$2yc_0 = 2c_1 + A. \tag{2.19}$$

This is a crucial point. As stated in the Introduction, it means that the associated wave function, $\chi_{\cos}(r,E) = \sum_n c_n(\mu)\phi_n(r)$, with these expansion coefficients does not solve the reference wave equation $(H_0 - E)|\chi\rangle = 0$ since $\sum_m J_{nm}c_m \neq 0$. However, Eq. (2.19) and the expression for $J_{nm}$ given by Eq. (2.6) imply that $\sum_m J_{nm}c_m = -\tfrac{\lambda A}{2}\left(\mu^2 + \tfrac{1}{4}\right)\delta_{n0}$. This means that $\chi_{\cos}$ solves the following regularized inhomogeneous wave equation

$$(H_0 - E)|\chi_{\cos}\rangle = -\tfrac{kA}{2\sin\theta}|\tilde{\phi}_0\rangle, \tag{2.20}$$

where, for $\ell = 0$, $\tilde{\phi}_0(r) = \lambda e^{-\lambda r/2}$ and $\langle\phi_n|\tilde{\phi}_0\rangle = \delta_{n0}$. It should be obvious that the right-hand side of this equation vanishes in the asymptotic region where $r \to \infty$. That is, asymptotically $\chi_{\cos}$ satisfies the same reference wave equation as does $\chi_{irr}$. Moreover, by equating the asymptotic behavior of $\chi_{\cos}(r,E)$ with that of $\chi_{irr}(r,E)$ one can easily find the value of $A$ to be $-\tfrac{1}{2}\sqrt{\pi}$ as explained in section III.

### B. The case $\alpha + \beta = \tfrac{1}{2}$

Again, maintaining positivity of $\beta$, this case gives two solutions as well. One is valid for all $\ell$ where Eq. (2.15) gives

$$\alpha = -\tfrac{1}{2}\ell,\ \beta = \tfrac{1}{2}(\ell+1),\ a = -n-\ell-\tfrac{1}{2},\ b = n+\ell+\tfrac{3}{2},\ \text{and}\ c = -\ell+\tfrac{1}{2}, \tag{2.21}$$

corresponding to
$$\left(\cos\tfrac{\theta}{2}\right)^{\ell+1}\left(\sin\tfrac{\theta}{2}\right)^{-\ell} {}_2F_1\!\left(-n-\ell-\tfrac{1}{2},n+\ell+\tfrac{3}{2};-\ell+\tfrac{1}{2};\sin^2\tfrac{\theta}{2}\right). \qquad (2.22)$$

This hypergeometric function is a non-terminating series because none of the first two arguments will ever be a negative integer. However, we can use the transformation [7],
$$_2F_1(a,b;c;z)=(1-z)^{c-a-b}{}_2F_1(c-a,c-b;c;z), \qquad (2.23)$$
to write it in the following alternative, but equivalent form
$$(\sin\theta)^{-\ell}{}_2F_1\!\left(-n-2\ell-1,n+1;-\ell+\tfrac{1}{2};\sin^2\tfrac{\theta}{2}\right). \qquad (2.24)$$

Now, this hypergeometric function is a finite polynomial, $P_n(E)$, of order $n+2\ell+1$ in $\sin^2\tfrac{\theta}{2}$. Using the fact that $c_n(E)=A_\ell\left[\Gamma(n+1)/\Gamma(n+2\ell+2)\right]P_n(E)$ we can therefore write
$$c_n(E)=A_\ell\tfrac{\Gamma(n+1)}{\Gamma(n+2\ell+2)}(\sin\theta)^{-\ell}{}_2F_1\!\left(-n-2\ell-1,n+1;-\ell+\tfrac{1}{2};\sin^2\tfrac{\theta}{2}\right), \qquad (2.25)$$
where $A_\ell$ is the normalization constant, which is independent of the energy $E$ and the index $n$. It is evaluated in the Appendix and obtained as $A_\ell=-2^{\ell-1}\Gamma\!\left(\ell+\tfrac{1}{2}\right)$. One can verify that this solution satisfies the three-term recursion relation (2.8) for all $\ell$ but not the initial relation (when $n = 0$). Instead, it satisfies the following inhomogeneous initial relation
$$2(\ell+1)y\,c_0 = 2(\ell+1)c_1 + (2\ell+1)A_\ell\big/\!\left[\Gamma(2\ell+2)(\sin\theta)^\ell\right]. \qquad (2.26)$$

The corresponding wavefunction does not satisfy the reference wave equation but an inhomogeneous one that reads
$$\left(H_0 - E\right)|\chi_{\cos}\rangle = -\left(\ell+\tfrac{1}{2}\right)A_\ell k(\sin\theta)^{-\ell-1}\big|\tilde\phi_0\big\rangle, \qquad (2.27)$$
where now, for all $\ell$, $\tilde\phi_0(r)=\tfrac{\lambda(\lambda r)^\ell}{\Gamma(2\ell+2)}e^{-\lambda r/2}$. It is easy to verify that the S-wave solution obtained above in (2.18) is a special case of (2.25) with $\ell = 0$ and $A = A_0$. Similar to the previous case, we also find another independent special solution for S-wave ($\ell = 0$) where,
$$\alpha=\tfrac{1}{2},\ \beta=0,\ a=-n-\tfrac{1}{2},\ b=n+\tfrac{3}{2},\ \text{and}\ c=\tfrac{3}{2}, \qquad (2.28)$$
corresponding to $(1-y)^{\frac{1}{2}}{}_2F_1\!\left(-n-\tfrac{1}{2},n+\tfrac{3}{2};\tfrac{3}{2};\tfrac{1-y}{2}\right)$. Using the transformation (2.23) this could be rewritten as $(\sin\theta){}_2F_1\!\left(-n,n+2;\tfrac{3}{2};\tfrac{1-y}{2}\right)$. Alternatively, we could write it as $\tfrac{\sin\theta}{n+1}U_n(y)$, where $U_n(y)$ is the Chebyshev polynomial of the second kind [7]. Therefore, the expansion coefficients of the reference wavefunction in (2.7) are
$$s_n(E)=\tfrac{B\sin\theta}{n+1}U_n(\cos\theta)=\tfrac{B}{n+1}\sin(n+1)\theta, \qquad (2.29)$$
where $B$ is a normalization constant, which is independent of the energy $E$ and the index $n$. Now, one can easily verify that this solution satisfies the three term recursion relation (2.8) with $\ell=0$, as well as its initial relation. That's why it was referred to as $s_n(E)$. In fact, one can easily show that this solution is a special case of that in (2.12) with $\ell=0$ and $B=2/\sqrt{\pi}$.

Now, we collect all findings and give a brief summary of the results obtained above for the 3D spherically symmetric problem with finite range potential, $V(r)$, and whose reference Hamiltonian, $H_0$, is the free kinetic energy operator. By relaxing the constraint on the matrix representation of $H_0$ from being diagonal to tridiagonal, a *natural* and special regularization of the reference problem emerges. We describe this regularization as *natural* due to the fact that neither an arbitrarily chosen constraint nor a regularizing parameter (e.g., a cut-of parameter) was introduced by hand. Two solutions are obtained as infinite expansion in the discrete square integrable basis (2.4). We found one of them to be identical to the regular solution of the problem where the expansion coefficients are given by (2.12). The other is a regularized version of the irregular reference solution with the expansion coefficients in (2.25). These regularized reference wavefunctions are used in scattering calculations by writing the asymptotic solution to the full problem, $H = H_0 + V$, as

$$\lim_{r \to \infty} \psi(r, E) = \chi_-(r, E) + e^{2i\delta(E)} \chi_+(r, E), \tag{2.30}$$

where $\chi_\pm(r, E) = \chi_{\cos}(r, E) \pm i \chi_{\sin}(r, E)$ and $\delta(E)$ is the energy-dependent phase shift that contains the contribution of the short range scattering potential $V(r)$. One can calculate $\delta(E)$ using any convenient approach based on the chosen scattering method [1,3]. In the following section we show that the regularization obtained above is equivalent to that already used in the J-matrix method of scattering.

### III. REGULARIZATION IN THE J-MATRIX METHOD

The J-matrix method is an algebraic method of quantum scattering with substantial success in atomic and nuclear physics [1]. Its structure in function space is endowed with formal and computational analogy to the R-matrix method in configuration space [3]. The method yields scattering information over a continuous range of energy for a model potential obtained by truncating the given short-range potential $V$ in a finite subset of the $L^2$ basis. It was extended to multi-channel [11] as well as relativistic scattering [12]. In the J-matrix method the irregular reference solution, $\chi_{irr}$, is replaced by a "regularized" solution, $\bar{\chi}_{irr}$, which is regular at the origin but asymptotically equals to $\chi_{irr}$. Thus, $\bar{\chi}_{irr}$ could be expanded in terms of the square integrable basis, $\{\phi_n\}_{n=0}^\infty$, as $|\bar{\chi}_{irr}\rangle = \sum_n a_n |\phi_n\rangle$. Obviously, $\bar{\chi}_{irr}$ does not satisfy the reference wave equation whereas $\chi_{irr}$ does. It is chosen to satisfy the inhomogeneous equation

$$(H_0 - E)\bar{\chi}_{irr}(r, E) = \beta \tilde{\tilde{\xi}}(r), \tag{3.1}$$

where $\beta$ is an energy dependent real parameter and $\tilde{\tilde{\xi}}(r)$ is a regularizing function that is chosen to belong to the space spanned by $\{\tilde{\phi}_n\}$. For a given $\tilde{\xi}(r)$ the parameter $\beta$ is evaluated by matching $\bar{\chi}_{irr}$ and $\chi_{irr}$ at the boundary of configuration space. The two-point Green's function $G_0(r, r', E)$ is formally defined as $G_0 = \langle r | (H_0 - E)^{-1} | r' \rangle$. Applying this on Eq. (3.1) we obtain [5]

$$\bar{\chi}_{irr}(r, E) = -\beta \int_0^\infty G_0(r, r', E) \tilde{\xi}(r') dr'. \tag{3.2}$$

Now, $G_0(r, r', E) = \frac{2}{W(E)} \chi_{reg}(r_<, E) \chi_{irr}(r_>, E)$, where $r_<(r_>)$ is the smaller (larger) of $r$ and $r'$ and $W(E)$ is the Wronskian of the two independent reference solutions $\chi_{reg}$ and $\chi_{irr}$ which is independent of $r$. Substituting this in Eq. (3.2) we get

$$\bar{\chi}_{irr}(r) = -\frac{2\beta}{W(E)}\left[\chi_{irr}(r)\int_0^r \chi_{reg}(r')\tilde{\xi}(r')dr' + \chi_{reg}(r)\int_r^\infty \chi_{irr}(r')\tilde{\xi}(r')dr'\right]. \quad (3.3)$$

Taking the limit as $r \to \infty$, where $\bar{\chi}_{irr}$ equals $\chi_{irr}$, we obtain

$$\beta(E) = -W(E)\Big/2\int_0^\infty \chi_{reg}(r',E)\tilde{\xi}(r')dr'. \quad (3.4)$$

For a given set of chosen regularization parameters, $\{b_n\}$, we can write $\tilde{\xi}(r) = \sum_n b_n \tilde{\phi}_n(r)$. Substituting this together with $\chi_{reg}(r,E) = \sum_n s_n(E)\phi_n(r)$ in Eq. (3.4) we obtain

$$\beta(E) = -W(E)\Big/2\sum_n b_n s_n(E). \quad (3.5)$$

In the classic version of the J-matrix method, regularization is performed by choosing the parameters $b_n = \delta_{n0}$. In that case, $\beta = -W/2s_0$ and Eq. (3.1) reads $(H_0 - E)\bar{\chi}_{irr}(r,E) = \beta\tilde{\phi}_0(r)$, which is identical to Eq. (2.27) with $\beta = -\left(\ell+\frac{1}{2}\right)A_\ell k(\sin\theta)^{-\ell-1}$. Therefore, regularization in the standard J-matrix method is equivalent to the natural choice obtained above.

## ACKNOWLEDGMENTS

Bahlouli and Abdelmonem acknowledge the support of King Fahd University of Petroleum and Minerals under project FT-2005/11. Al-Ameen and Al-Abdulaal are grateful to Girls College of Sciences, higher studies section, and the physics department for their support.

## APPENDIX A
## CALCULATING THE NORMALIZATION CONSTANT $A_\ell$ IN EQ. (2.25)

There might be several ways to obtain the normalization coefficient $A_\ell$. A direct approach is to equate the asymptotic expression of $\chi_{irr}(r,E)$ in Eq. (2.3b) to that of $\chi_{\cos}(r,E)$ with the expansion coefficients given by Eq. (2.25). In such an approach, one utilizes the asymptotic behavior of the Laguerre polynomials [7]. However, a simpler approach is to use the Green's function method outlined in Sec. III. It goes as follows. By comparing Eq. (3.1) to Eq. (2.27) we obtain $\beta = -\left(\ell+\frac{1}{2}\right)A_\ell k(\sin\theta)^{-\ell-1}$ which when inserted in Eq. (3.4) gives $\beta = -W/2\langle\chi_{reg}|\tilde{\phi}_0\rangle$. That is, $\beta(E) = -W(E)/2s_0(E)$. Now, for our problem, which is defined by the reference wave equation (2.2) and solutions (2.3), the Wronskian is $-k$. Using this and the value of $s_0(E)$ given by Eq (2.12) for $n = 0$ along with the fact that $\Gamma(2\ell+2) = \frac{1}{\sqrt{\pi}}2^{2\ell+1}\Gamma(\ell+1)\Gamma\left(\ell+\frac{3}{2}\right)$ we obtain $A_\ell = -2^{\ell-1}\Gamma\left(\ell+\frac{1}{2}\right)$.